\documentclass{article}
\usepackage{graphicx}
\usepackage{amsmath}
\usepackage{amsfonts}
\usepackage{amssymb}

\begin{document}

\title{Square root voting system,\\optimal threshold and $\pi$}
\author{Karol \.{Z}yczkowski$^{a,b}$ and Wojciech S\l omczy\'{n}ski$^{c}$\\$^{a}$Institute of Physics, Jagiellonian University, \\ul. Reymonta 4, 30-059 Krak\'{o}w, Poland \\$^{b}$Center for Theoretical Physics, Polish Academy of Sciences, \\Al. Lotnik\'{o}w 32/46, 02-668 Warszawa, Poland\\$^{c}$Institute of Mathematics, Jagiellonian University, \\ul. \L ojasiewicza 6, 30-348 Krak\'{o}w, Poland}
\date{March 26, 2012}
\maketitle

Abstract:

The problem of designing an optimal weighted voting system for the two-tier
voting, applicable in the case of the Council of Ministers of the European
Union (EU), is investigated. Various arguments in favor of the square root
voting system, where the voting weights of member states are proportional to
the square root of their population are discussed and a link between this
solution and the random walk in the one-dimensional lattice is established. It
is known that the voting power of every member state is approximately equal to
its voting weight, if the threshold $q$ for the qualified majority in the
voting body is optimally chosen. We analyze the square root voting system for
a generic `union' of $M$ states and derive in this case an explicit
approximate formula for the level of the optimal threshold: $q\simeq
1/2+1/\sqrt{\pi M}$. The prefactor $1/\sqrt{\pi}$ appears here as a result of
averaging over the ensemble of `unions' with random populations.

\medskip

Keywords:

Banzhaf index, Penrose law, optimal threshold, random states

\bigskip$^{*}$This paper was presented to the \textsl{Voting Power in Practice
Symposium} at the London School of Economics, 20-22 March 2011, sponsored by
the Leverhulme Trust.

\bigskip e-mail: karol@tatry.if.uj.edu.pl \quad\quad wojciech.slomczynski@im.uj.edu.pl

\medskip\newpage

\section{Introduction}

Recent political debate on the voting system used in the Council of Ministers
of the European Union stimulated research in the theory of indirect voting,
see e.g. \cite{FelMac01,Lee02,ACL03,PajWid04,BeiBovHar05}. The \textsl{double
majority} voting system, adopted for the Council by The Treaty of Lisbon in
December 2007 is based on two criteria: `per capita' and `per state'. This
system apparently reflects the principles of equality of Member States and
that of equality of citizens. However, as recently analyzed by various authors
\cite{BalWid04,Ade06,SloZycAPP06,ABF07,Hos08,BarTac09,Kir10,Mob10,LeeAzi10,Puk10,SloZyc10}%
, in such a system the large states gain a lot of power from the direct link
to population, while the smallest states derive disproportionate power from
the other criterion. The combined effect saps influence away from all
medium-sized countries. Ironically, a similar conclusion follows from a book
by Lionel Penrose, who wrote already in 1952 \cite{Pen52}:

\textsl{If two votings were required for every decision, one on a per capita
basis and the other upon the basis of a single vote for each country, this
system would be inaccurate in that it would tend to favor large countries.}

To quantify the notion of voting power, mathematicians introduced the concept
of power index of a member of the voting body, which measures the probability
that his vote will be decisive in a hypothetical ballot: Should this member
decide to change his vote, the winning coalition would fail to satisfy the
qualified majority condition. Without any further information about the voting
body it is natural to assume that all potential coalitions are equally likely.
This very assumption leads to the concept of Banzhaf(-Penrose) index called so
after John Banzhaf, an American attorney, who introduced this index
independently in 1965 \cite{Ban65}.

Note that this approach is purely normative, not descriptive: we are
interested in the potential voting power arising from the voting procedure
itself. Calculation of the voting power based on the counting of majority
coalitions is applicable while analyzing institutions in which alliances are
not permanent, but change depending upon the nature of the matter under consideration.

To design a representative voting system, i.e. the system based on the
democratic principle, that the vote of any citizen of any Member State is of
equal worth, one needs to use a weighted voting system. Consider elections of
the government in a state with population of size $N$. It is easy to imagine
that an average German citizen has smaller influence on the election of his
government than, for example, a citizen of the neighboring Luxembourg.
Analyzing this problem in the context of voting in the United Nations just
after the World War II Penrose showed, under some natural assumptions, that in
such elections the voting power of a single citizen decays as one over square
root of $N$. Thus, the system of indirect voting applied to the Council is
representative, if the voting power of each country is proportional to the
square root of $N$, so that both factors cancel out. This statement is known
in the literature under the name of the \textsl{Penrose square root law}
\cite{Pen46,FelMac98}. It implies that the voting power of each member of the
EU Council should behave as $\sqrt{N}$ and such voting systems have been
analyzed in this context by several experts since late 90s
\cite{FelMac97,LarWid98}.

It is challenging to explain this fact in a way accessible to a wide audience
\cite{ZycSloZas06,KSZ07,Puk07,Pop07}. A slightly paradoxical nonlinearity in
the result of Penrose is due to the fact that voting in the Council should be
considered as a two--tier voting system: Each member state elects a
government, which delegates its representative to the Council. Any
representative has to say \textsl{`Yes'} or \textsl{`No'} on behalf of his
state in every voting organized in the Council. The key point is that in such
a voting each member of the Council cannot split his vote. Making an
idealistic assumption that the vote of a Minster in the Council represents the
will of the majority of the citizens of the state he represents, his vote
`\textsl{Yes}' means only that a majority of the population of his state
supports this decision, but does not reflect the presence of a minority.

Consider an exemplary issue to be voted in the Council and assume that the
preferences of the voters in each state are known. Assume hypothetically that
a majority of population of Malta says `\textsl{Yes}' on a certain issue, the
votes in Italy split as $30$ millions `\textsl{Yes}' and $29$ millions
`\textsl{No}', while all $43$ millions of citizens of Spain say `\textsl{No}'.
A member of the Council from Malta follows the will of the majority
in his state and votes `\textsl{Yes}'. So does the representative of Italy.
According to the double majority voting system his vote is counted on behalf
of the total number of $59$ millions of the population of Italy. Thus these
voting rules allow $30$ millions of voters in Italy to over-vote not only the
minority of $29$ millions in their state (which is fine), but also, with the
help of less than half a million of people from Malta, to over--vote $43$
millions of Spaniards.

This pedagogical example allows one to conclude that the double majority
voting system would work perfectly, if all voters in each member state had the
same opinion on every issue. Obviously such an assumption is not realistic,
especially in the case of the European states, in which the citizens can
nowadays afford the luxury of an independent point of view. In general, if a
member of the Council votes `\textsl{Yes}' on a certain issue, in an ideal
case one may assume that the number of the citizens of his state which support
this decision varies from $50\%$ till $100\%$ of the total population. In
practice, no concrete numbers for each state are known, so to estimate the
total number of European citizens supporting a given decision of the Council
one has to rely on statistical reasoning.

To construct the voting system in the Council with voting powers proportional
to the square root of populations one can consider the situation, where voting
weights are proportional to the square root of populations and the Council
takes its decision according to the principle of a \textsl{qualified
majority}. In other words, the voting in the Council yields acceptance, if the
sum of the voting weights of all Ministers voting `\textsl{Yes}' exceeds a
fixed quota $q$, set for the qualified majority. From this perspective the
quota $q$ can be treated as a free parameter \cite{LeeMac03,Mac10}, which may
be optimized in such a way that the mean discrepancy $\Delta$ between the
voting power (measured by the Banzhaf index) and the voting weight of each
member state is minimal.

In the case of the population in the EU consisting of $25$ member states it
was shown \cite{SloZyc04,ZycSloZas06} that the value of the optimal quota
$q_{\ast}$ for qualified majority in the Penrose's square root system is equal
to $62.0\%$, while for EU-$27$ this number drops down to $61.5\%$
\cite{SloZycAPP06,SloZyc07}. Furthermore, the optimal quota can be called
\textsl{critical}, since in this case the mean discrepancy $\Delta(q_{\ast})$
is very close to zero and thus the voting power of every citizen in each
member state of the Union is practically equal. This simple scheme of voting
in the EU Council based on the square root law of Penrose supplemented by a
rule setting the optimal quota to $q_{\ast}$ happens to give larger voting
powers to the largest EU than the Treaty of Nice, but smaller ones than the
Treaty of Lisbon. Therefore this voting system has been dubbed by the media as
the \textsl{Jagiellonian Compromise}.

It is known that the existence of the critical quota $q_{\ast}$, is not
restricted to the particular distribution of the population in the European
Union, but it is also characteristic of a generic distribution of the
population \cite{SloZyc04,ChaChuMac06,SloZycAPP06}. The value of the critical
quota depends on the particular distribution of the population in the `union',
but even more importantly, it varies considerably with the number $M$ of member
states. An explicit approximate formula for the critical quota was derived in
\cite{SloZyc07}. It is valid in the case of a relatively large number of the members of
the `union' and in the asymptotic limit, $M\rightarrow\infty$, the critical
quota tends to $50\%$, in consistence with the so-called \emph{Penrose limit theorem}
\cite{LinMac04}.

On one hand it is straightforward to apply this explicit formula for the
current population of all member states of the existing European Union, as
well as to take into account various possible scenarios of a possible
extension of the Union. On the other hand, if the number of member states is
fixed, while their populations vary in time, continuous update of the optimal
value for the qualified majority may be cumbersome and unpractical. Hence one
may try to neglect the dependence on the particular distribution of the
population by selecting for the quota the mean value of $\langle q\rangle$,
where the average is taken over a sample of random population distributions,
distributed uniformly in the allowed space of $M$-point probability
distributions. In this work we perform such a task and derive an explicit,
though approximate, formula for the average critical quota.

This paper is organized as follows. In section 2 devoted to the one-tier
voting system, we recall the definition of Banzhaf index and review the
Penrose square root law. In section 3, which concerns the two-tier voting
systems, we describe the square root voting system and analyze the average
number of misrepresented voters. Section 4 is devoted to the problem of
finding the optimal quota for the qualified majority. It contains the key
result of this paper: derivation of a simple approximate formula for the
average optimal quota, which depends only on the number $M$ of the member
states and is obtained by averaging over an ensemble of random distributions
of the population of the `union'.

\section{One tier voting}

Consider a voting body consisting of $M$ voters voting according to the
qualified majority rule. Assume that the weights of the votes need not to be
equal, which is typical e.g. in the case of an assembly of stockholders of a
company: the weight of the vote of a stockholder depends on the number of
shares he or she possesses. It is worth to stress that, generally, the voting
weights do not directly give the voting power.

To quantify the a priori voting power of any member of a given voting body
game theorists introduced the notion of a power index. It measures the
probability that a member's vote will be decisive in a hypothetical ballot:
should this player decide to change its vote, the winning coalition would fail
to satisfy the qualified majority condition. In the game theory approach to
voting such a player is called \textsl{pivotal}.

The assumption that all potential coalitions of voters are equally likely
leads to the concept of the Banzhaf index \cite{Pen46,Ban65}. To compute this
power index for a concrete case one needs to enumerate all possible
coalitions, identify all winning coalitions, and for each player find the
number of cases in which his vote is decisive.

Let $M$ denote the number of voters and $\omega$ the total number of all
winning coalitions, that satisfy the qualified majority condition. Assume that
$\omega_{k}$ denotes the number of winning coalitions that include the $k$-th
player; where $k=1,\ldots,M$. Then the Banzhaf index of the $k$--th voter
reads
\begin{equation}
\psi_{k}\ :=\ \frac{\omega_{k}-(\omega-\omega_{k})}{2^{M-1}}=\frac{2\omega
_{k}-\omega}{2^{M-1}}\ .\label{PB1}%
\end{equation}
To compare these indices for decision bodies consisting of different number of
players, it is convenient to define the \textsl{normalized Banzhaf (-Penrose)
index}:
\begin{equation}
\beta_{k}\ :=\ \frac{\psi_{k}}{\sum_{i=1}^{M}\psi_{i}}\,\label{PB2}%
\end{equation}
such that $\sum_{i=1}^{M}\beta_{i}=1$.

In the case of a small voting body such a calculation is straightforward,
while for a larger number of voters one has to use a suitable computer program.

\subsection{Square root law of Penrose\label{sec:srl}}

Consider now the case of $N$ members of the voting body, each given a single
vote. Assume that the body votes according to the standard majority rule. On
one hand, since the weights of each voter are equal, so must be their voting
powers. On the other hand, we may ask, what happens if the size $N$ of the
voting body changes, for instance, if the number of eligible voters gets
doubled, how does this fact influence the voting power of each voter?

For simplicity assume for a while that the number of voters is odd, $N=2j+1$.
Following original arguments of Penrose we conclude that a given voter will be
able to effectively influence the outcome of the voting only if the votes
split half and half: If the vote of $j$ players would be `\textsl{Yes}' while
the remaining $j$ players vote `\textsl{No}', the role of the voter we analyze
will be decisive.

Basing upon the assumption that all coalitions are equally likely one can ask,
how often such a case will occur? In mathematical language the model in which
this assumption is satisfied is equivalent to the \textsl{Bernoulli scheme}.
The probability that out of $2j$ independent trials we obtain $k$ successes
reads
\begin{equation}
P_{k}:=\ \binom{2j}{k}p^{k}(1-p)^{2j-k}\ \text{,}\label{Bern1}%
\end{equation}
where $p$ denotes the probability of success in each event. In the simplest
symmetric case we set $p=1-p=1/2$ and obtain
\begin{equation}
P_{j}\ =\ \left(  \frac{1}{2}\right)  ^{2j}\frac{\left(  2j\right)  !}{\left(
j!\right)  ^{2}}\ \text{.}\label{Bern2}%
\end{equation}
For large $N$ we may use the Stirling approximation for the factorial and
obtain the probability $\psi$ that the vote of a given voter is decisive
\begin{equation}
\psi=P_{j}\ \sim\ 2^{-2j}\frac{(2j/e)^{2j}\sqrt{4\pi j}}{[(j/e)^{j}\sqrt{2\pi
j}]^{2}}\ =\ \frac{1}{\sqrt{\pi j}}\ \sim\ \sqrt{\frac{2}{\pi N}}%
\ \text{.}\label{Bern3}%
\end{equation}
For $N$ even we get the same approximation. In this way one can show that the
voting power of any member of the voting body depends on its size as
$1/\sqrt{N}$, which is the \textsl{Penrose square root law}. The above result
is obtained under the assumption that the votes of all citizens are
uncorrelated. A sound mathematical investigation of the influence of possible
correlations between the voting behavior of individual citizens for their
voting power has been recently presented by Kirsch \cite{Kir07}. It is easy to
see that due to strong correlations certain deviations from the square root
law have to occur, since in the limiting case of unanimous voting in each
state (perfect correlations), the voting power of a single citizen from a
state with population $N$ will be inversely proportional to $N$.

The issue that the assumptions leading to the Penrose law are not exactly
satisfied in reality was raised many times in the literature, see, e.g.
\cite{GKT02,GKB04}, also in the context of the voting in the Council of the
European Union \cite{LarVal08}. However, it seems not to be easy to design a
rival model voting system which correctly takes into account the essential
correlations, varying from case to case and evolving in time. Furthermore, it
was argued \cite{Kir07} that the strength of the correlations between the
voters tend to decrease in time. Thus, if one is to design a voting system to
be used in the future in the Council of the European Union, it is reasonable
to consider the idealistic case of no correlations between individual voters.
We will follow this strategy and in the sequel rely on the square root law of Penrose.

\subsection{Pivotal voter and the return probability in a random walk}

It is worth to emphasize that the square root function appearing in the above
derivation is typical to several other reasonings in mathematics, statistics
and physics. For instance, in the analyzed case of a large voting body, the
probability distribution $P_{k}$ in the Bernoulli scheme can be approximated
by the Gaussian distribution with the standard deviation being proportional to
$1/\sqrt{N}$. It is also instructive to compare the above voting problem with
a simple model of a random walk on the one dimensional lattice.

Assume that a particle subject to external influences in each step jumps a
unit distance left or right with probability one half. What is the probability
that it returns to the initial position after $N$ steps? It is easy to see
that the probability scales as $1/\sqrt{N}$, since the answer is provided by
exactly the same reasoning as for the Penrose law.

Consider an ensemble of particles localized initially at the zero point and
performing such a random walk on the lattice. If the position of a particle at
time $n$ differs from zero, in half of all cases it will jump towards zero,
while in the remaining half of cases it will move in the opposite direction.
Hence the \textsl{mean} distance $\langle D\rangle$ of the particle from zero
will not change. On the other hand, if at time $n$ the particle happened to
return to the initial position, in the next step it would certainly jump away
from it, so the mean distance from zero would increase by one.

To compute the mean distance form zero for an ensemble of random particles
performing $N$ steps, we need to sum over all the cases, when the particle
returns to the initial point. Making use of the previous result, that the
return probability $P(n)$ at time $n$ behaves as $1/\sqrt{n}$, we infer that
during the time $N$ the mean distance behaves as
\begin{equation}
\langle D(N)\rangle\ \approx\ \sum_{n=1}^{N}P(n)\ \approx\ \sum_{n=1}^{N}%
\frac{1}{\sqrt{n}}\ \sim\ \sqrt{N}\ \text{.}\label{diff}%
\end{equation}
This is just one formulation of the \textsl{diffusion law}. As shown, the
square root of Penrose is closely related with some well known results from
mathematics and physics, including the Gaussian approximation of binomial
distribution and the diffusion law.

\section{Two tier voting}

In a two-tier voting system each voter has the right to elect his
representative, who votes on his behalf in the upper chamber. The key
assumption is that, on one hand, he should represent the will of the
population of his state as best he can, but, on the other hand, he is obliged
to vote `\textsl{Yes}' or `\textsl{No}' in each ballot and cannot split his
vote. This is just the case of voting in the Council of the EU, since citizens
in each member state choose their government, which sends its Minister to
represent the entire state in the Council.

These days one uses in the Council the triple majority system adopted in 2001
in the \textsl{Treaty of Nice}. The Treaty assigned to each state a certain
number of `weights', distributed in an ad hoc fashion. The decision of the
Council is taken if the coalition voting in favour of it satisfies three conditions:

a) it is formed by the standard majority of the member states,

b) states forming the coalition represent more then $62\%$ of the entire
population of the Union,

c) the total number of weights of the `\textsl{Yes}' votes exceeds a quota
equal to approximately $73.9\%$ of all weights.

Although all three requirements have to be fulfilled simultaneously, detailed
analysis shows that condition c) plays a decisive role in this case: if it is
satisfied, the two others will be satisfied with a great likelihood as well
\cite{FelMac01,Lee02}.

Therefore, the voting weights in the Nice system play a crucial role. However,
the experts agree \cite{FelMac01,PajWid04} that the choice of the weights
adopted is far from being optimal. For instance the voting power of some
states (including e.g. Germany and Romania) is significantly smaller than in
the square root system. This observation is consistent with the fact that
Germany was directly interested to abandon the Nice system and push toward
another solution that would shift the balance of power in favor of the
largest states.

In the double majority voting system, adopted in December 2007 in Lisbon, one
gave up the voting weights used to specify the requirement c) and decided to
preserve the remaining two conditions with modified majority quotas. A
coalition is winning if:

a') it is formed by at least $55\%$ of the members states,

b') it represents at least $65\%$ of the population of the Union.

Additionally, every coalition consisting of all but three (or less) countries
is winning even if it represents less than $65\%$ of the population of the Union.

The double majority system will be used in the Council starting from the year
2014. However, a detailed analysis by Moberg \cite{Mob10} shows that in this
concrete case the `double majority' system is not really double, as the per
capita criterion b') plays the dominant role here. In comparison with the
Treaty of Nice, the voting power index will increase for the four largest
states of the Union (Germany, France, the United Kingdom and Italy) and also
for the smallest states. To understand this effect we shall analyze the voting
system in which the voting weight of a given state is directly proportional to
its population.

\subsection{Voting systems with per capita criterion}

The idea `one citizen -- one vote' looks so natural and appealing, that in
several political debates one often did not care to analyze in detail its
assumptions and all its consequences. It is somehow obvious that a minister
representing a larger (if population is considered) state should have a larger
weight during each voting in the EU Council. On the other hand, one needs to
examine whether the voting weights of a minister in the Council should be
proportional to the population he represents. It is clear that this would be
very much the case, if one could assume that all citizens in each member state
share the very same opinion in each case.

However, this assumption is obviously false, and nowadays we enjoy in Europe
the freedom to express various opinions on every issue. Let us then formulate
the question, how many citizens from his state each minister actually
represents in an exemplary voting in the Council? Or to be more precise, how
many voters from a given state with population $N$ share in a certain case the
opinion of their representative? We do not know!

Under the idealizing assumption that the minister always votes according to
the will of the majority of citizens in his state, the answer can vary from
$N/2$ to $N$. Therefore, the difference between the number of the citizens
supporting the vote of their minister and the number of those who are against
it can vary from $0$ to $N$. In fact it will vary from case to case in this
range, so an assumption that it is always proportional to $N$ is false. This
crucial issue, often overlooked in popular debates, causes problems with
representativeness of a voting system based on the `per capita' criterion.

There is no better way to tackle the problem as to rely on certain statistical
assumptions and estimate the \textsl{average} number of `satisfied citizens'.
As such an analysis is performed later in this paper, we shall review here
various arguments showing that a system with voting weights directly
proportional to the population is advantageous to the largest states of the union.

Consider first a realistic example of a union of nine states: a large state
$A$, with $80$ millions of citizens and eight small states from $B$ to $I$,
with $10$ millions each. Assume now that in a certain case the distribution of
the opinion in the entire union is exactly polarized: in each state
approximately $50\%$ of the population support the vote `\textsl{Yes}', while
the other half is against. Assume now that the government of the large state
is in position to establish exactly the will of the majority of citizens in
their state (say it is the vote `\textsl{Yes}') and order its minister to vote
accordingly. Thus the vote of this minister in the council will then be
counted as a vote of $80$ millions of citizens.

On the other hand, in the remaining states the probability that the majority
of citizens support `\textsl{Yes}' is close to $50\%$. Hence it is most likely
that the votes of the ministers from the smaller states split as $4:4$. Other
outcomes: $5:3$, $6:2$, or $7:1$ are less probable, but all of them result in
the majority of the representative of the large state $A$. The outcome $8:0$
is much less likely, so if we sum the votes of all nine ministers we see that
the vote of the minister from the largest state will be decisive. Hence we
have shown that the voting power of all citizens of the nine small states is
negligible, and the decision for this model union is practically taken by the
half of its population belonging to the largest state $A$. Even though in this
example we concentrated on the `per capita' criterion and did not take into
account the other criterion, it is not difficult to come up with analogous
examples which show that the largest states are privileged also in the double
majority system. Similarly, the smallest states of the union benefit from the
`per state' criterion.

Let us have a look at the position of the minority in large states. In the
above example the minority in the $80$ million state can be as large as $40$
million citizens, but their opinion will not influence the outcome of the
voting, independently of the polarization of opinion in the remaining eight
states. Thus one may conclude that in the voting system based on the `per
capita' criterion, the influence of the politicians representing the majority
in a large state is enhanced at the expense of the minority in this state and
the politicians representing the smaller states.

Last but not least, let us compare the maximal sizes of the minority, which
can arise during any voting in an EU member state. In Luxembourg, with its
population of about $400\,000$ people, the minority cannot exceed $200\,000$
citizens. On the other hand, in Germany, which is a much larger country, it is
possible that the minority exceeds $41$ millions of citizens, since the total
population exceeds $82$ millions. It is then fair to say, that, due to
elections in smaller states, we know the opinion of citizens in these states
with a better accuracy, than in larger members of the union. Thus, as in
smaller states the number of misrepresented citizens is smaller, their votes
in the EU Council should be weighted by larger weights than the vote of the
largest states. This very idea is realized in the weighted voting system
advocated by Penrose.

\subsection{Square root voting system of Penrose}

The Penrose system for the two-tier voting is based on the square root law
reviewed in Sec. \ref{sec:srl}. Since the voting power of a citizen in state
$k$ with population $N_{k}$ scales as $1/\sqrt{N_{k}}$, this factor will be
compensated, if the voting power of each representative in the upper chamber
will behave as $\sqrt{N_{k}}$. Only in this way the voting power of each
citizen in every state of a union consisting of $M$ states will be equal.

Although we know that the voting power of a minister in the Council needs not
coincide with the weight of his vote, as a rough approximation let us put his
weights $w_{k}$ proportional to the square root of the population he
represents, that is $w_{k}=\sqrt{N_{k}}/\sum_{i=1}^{M}\sqrt{N_{i}}$.

To see a possible impact of the change of the weights let us now return to the
previous example of a union of one big state and eight small ones. As the
state $A$ is $8$ times as large as each of the remaining states, its weight in
the Penrose system will be $w_{A}=\sqrt{8}w_{B}$. As $\sqrt{8}$ exceeds $2$
and is smaller then $3$, we see that accepting the Penrose system will
increase the role of the minority in the large state and the voting power of
all smaller states. For instance, if the large state votes `\textsl{Yes}' and
the votes in the eight states split as $2:6$ or $1:7$ in favor for
`\textsl{No}', the decision will not be taken by the council, in contrast to
the simple system with one `per capita' criterion. There, we have assumed that
the standard majority of weights is sufficient to form a winning coalition. If
the threshold for the qualified majority is increased to $54\%$, also the
outcome $3:5$ in favor for `\textsl{No}' in the smaller states suffices to
block the decision taken in the large state.

This simple example shows that varying the quota for the qualified majority
considerably influences the voting power, see also \cite{LeeMac03,Mac10}. The
issue of the selection of the optimal quota will be analyzed in detail in the
subsequent section. At this point, it is sufficient to add that in general it
is possible to find such a level of the quota for which the voting power
$\beta_{k}$ of the $k$-th state is proportional to $\sqrt{N_{k}}$ and, in
consequence, the Penrose law is almost exactly fulfilled
\cite{SloZyc04,SloZycAPP06}.

Applying the square root voting system of Penrose combined with the optimal
quota to the problem of the Council, one obtains a fair solution, in which
every citizen in each member state of the Union has the same voting power,
hence the same influence on the decisions taken by the Council. In this case,
the voting power of each European state measured by the Banzhaf index scales
as the square root of its population. This weighted voting system happens to
give a larger voting power to the largest EU states (including Germany) than
the Treaty of Nice but smaller than the double majority system. On the other
hand, this system is more favorable to all middle size states then the double
majority, so it is fair to consider it as a compromise solution. The square
root voting system of Penrose is simple (one criterion only), transparent and
efficient -- the probability of forming a winning coalition is reasonably
high. Furthermore, as discussed later, it can be easily adopted to any
possible extension of the Union.

\subsection{The second square root law of Morris}

To provide an additional argument in favour of the square root weights of
Penrose \cite{FelMac99}, consider a model state of $N$ citizens, of which a
certain number $k$ support a given legislation to be voted in the council.
Assume that the representative of this state knows the opinion of his people
and, according to the will of the majority, he votes `\textsl{Yes}' in the
council if $k\geq N/2$. Then the number of citizens satisfied with his
decision is $k$. The number $N-k$ of disappointed citizens compensates the
same number of yes--votes, so the vote of the minister should effectively
represent the \textsl{difference} between them, $w=k-(N-k)=2k-N$. By our
assumption concerning the majority this number is positive, but in general the
effective weight of the vote of the representative should be $w=|2k-N|$.

Assume now that the votes of any of $N$ citizens of the state are independent,
and that both decisions are equally likely, so that $p=1-p=1/2$. Thus, for the
statistical analysis, we can use the Bernoulli scheme (\ref{Bern1}) and
estimate the weight of the vote of the minister by the average using the
Stirling approximation:
\begin{align}
\langle w_{N}\rangle\  &  =\ \sum_{k=0}^{N}P_{k}|2k-N|\ =\ \sum_{k=0}%
^{N}\binom{N}{k}\frac{1}{2^{N}}\left|  2k-N\right|  \ \nonumber\\
&  =\frac{\left\lfloor N/2\right\rfloor +1}{2^{N-1}}\binom{N}{\left\lfloor
N/2\right\rfloor +1}\ \sim\ \sqrt{\frac{2N}{\pi}}.
\end{align}
Here $\lfloor x\rfloor$ denotes the largest integer not greater than $x$. This
result provides another argument in favor of the weighted voting system of
Penrose: Counting all citizens of a given state, we would attribute the
weights of the representative proportionally to the population $N$ he is
supposed to represent. On the other hand, if we take into account the obvious
fact that not all citizens in this state share the opinion of the government
on a concrete issue and consider the average number of the majority of
citizens which support his decision one should weight his vote proportionally
to $\sqrt{N}$. From this fact one can deduce the \textsl{second square root
law of Morriss} \cite{Mor87,FelMac98,FelMac99,LarVal08} that states that the
average number of misrepresented voters in the union is smallest if the
weights are proportional to the square root of the population and quota is
equal to $50\%$, provided that the population of each member state is large
enough. Simultaneously, in this situation, the total voting power of the union
measured by the sum of the Banzhaf indices of all citizens in the union is maximal.

To illustrate the result consider a model union consisting of one large state
with population of $49$ millions, three medium states with $16$ million each
and three small with $1$ million citizens. For simplicity assume that the
double majority system and the Penrose system are based on the standard
majority of $50\%$. If the polarization of opinion in each state on a given
issue is as in the table below, only $39\%$ of the population of the union is
in favor of the legislative. However, under the rules of the double majority
system the decision is taken (against the will of the vast majority!), what is
not the case in the Penrose system, for which the coalition gains only $10$
votes out of $22$, so it fails to gather the required quota.\medskip

\begin{center}
\begin{table}[th]
\label{tab1} {\renewcommand{\arraystretch}{1.45}
\begin{tabular}
[c]{||l||c|c|c|c|c|c|c||l|c||}\hline\hline
State & $A$ & $B$ & $C$ & $D$ & $E$ & $F$ & $G$ & Total & \\
Population [M] & $1$ & $1$ & $1$ & $16$ & $16$ & $16$ & $49$ & $100$ & \\
Votes: Yes [M] & $2/3$ & $2/3$ & $2/3$ & $4$ & $4$ & $4$ & $25$ & $39$ & \\
Votes: No [M] & $1/3$ & $1/3$ & $1/3$ & $12$ & $12$ & $12$ & $24$ & $61$ & \\
State votes & $1$ & $1$ & $1$ & $0$ & $0$ & $0$ & $1$ & $4/7$ & \text{Y}\\
Minister's votes & $1$ & $1$ & $1$ & $0$ & $0$ & $0$ & $49$ & $52/100$ &
\text{Y}\\
Square root weights & $1$ & $1$ & $1$ & $4$ & $4$ & $4$ & $7$ & $22$ & \\
Square root votes & $1$ & $1$ & $1$ & $0$ & $0$ & $0$ & $7$ & $10/22$ &
\text{N}\\\hline\hline
\end{tabular}
}\end{table}
\end{center}

Table 1. Case study: Voting in the council of a model union of 7 members under
a hypothetical distribution of population and voting preferences. Although
$61\%$ of the total population of the union is against a legislative it will
be taken by the council, if the rules of the double majority are used. The
outcome of the voting according to the weighted voting system of Penrose
correctly reflects the will of the majority in the union.\medskip

To qualitatively understand this result, consider the minister representing
the largest country $G$ with a population of $49$ millions. In the double
majority system he uses his $49$ votes against the will of $24$ millions of
inhabitants. By contrast, the minister of the small state $A$ will
misrepresent at most one half of the million of his compatriots. In other
words, the precision in determining the will of all the citizens is largest in
the smaller states, so the vote of their ministers should gain a higher weight
than proportional to population, which is the case in the Penrose system.

\section{Optimal quota for qualified majority}

Designing a voting system for the Council one needs to set the threshold for
the qualified majority. In general, this quota can be treated as a free
parameter of the system and is often considered as a number to be negotiated.
For political reasons one usually requires that the voting system should be
\textsl{moderately conservative}, so one considers the quota in the wide range
from $55\%$ to $75\%$.

However, designing the voting system based on the theory of Penrose, one can
find a way to obtain a single number as the optimal value of the quota. In
order to assure that the voting powers of all citizens in the `union' are
equal one has to impose the requirement that the voting power of each member
state should be proportional to the square root of the population of each state.

Let us analyze the problem of $M$ members of the voting body, each
representing a state with population $N_{i}$, $i=1,\dots,M$. Denote by $w_{i}$
the voting weight attributed to each representative. We work with renormalized
quantities, such that $\sum_{i=1}^{M}w_{i}=1$. Assume that the decision of the
voting body is taken, if the sum of the weights $w_{i}$ of all members of the
coalition exceeds the given quota $q$.

In the Penrose voting system one sets the voting weights proportional to the
square root of the population of each state, $w_{i}\sim\sqrt{N_{i}}$ for
$i=1,\dots,M$. For any level of the quota $q$ one may compute numerically the
power indices $\beta_{i}$. To characterize the overall representativeness of
the voting system one may use various indices designed to quantify the
resulting inequality in the distribution of power among citizens \cite{LV02}.
Analyzing the influence of the quota $q$ for the average inequality of the
voting power we are going to use the mean discrepancy $\Delta$, defined as:%
\begin{equation}
\Delta:=\sqrt{\frac{1}{M}\sum_{i=1}^{M}(\beta_{i}-w_{i})^{2}}\ \text{,}
\label{Delta}%
\end{equation}
If the discrepancy $\Delta$ is equal to zero, the voting power of each state
is proportional to the square root of its population. Under the assumption
that the Penrose law is fulfilled, in such a case the voting power of any
citizen in each state is the same.

In practice, the coefficient $\Delta$ will not be exactly equal to zero, but
one may try to minimize this quantity. The optimal quota $q_{\ast}$ can be
defined as the quota for which the discrepancy $\Delta$ is minimal. Let us
note, however, that this definition works fine for the Banzhaf index, while
the dependence of the Shapley--Shubik index \cite{ShaShu54} on the quota does
not exhibit such a minimum.

Studying the problem for a concrete distribution of the population in the
European Union, it was found \cite{SloZyc04} that in these cases all $M$
ratios $\beta_{i}/w_{i}$ for $i=1,\dots,M$, plotted as a function of the quota
$q$, cross approximately near a single point. In other words, the discrepancy
$\Delta$ at this critical point $q_{\ast}$ is negligible. Numerical analysis
allows one to conclude that this optimal quota is approximately equal to
$62.0\%$ for the EU-$25$ \cite{SloZyc04}. At this very level of the quota the
voting system can be considered as optimal, since the voting power of all
citizens becomes equal. Performing detailed calculations one needs to care to
approximate the square root function with a sufficient accuracy, since the
rounding effects may play a significant role \cite{Kur07}.

It is worth to emphasize that in general the value of the optimal quota
decreases with the number of member states. For instance, in the case of the
EU-$27$ is is equal to $61.5\%$ \cite{ZycSloZas06,SloZyc07}, see Table 2. The
optimal quota was also found for other voting bodies including various
scenarios for an EU enlargement -- see Leech and Aziz \cite{LeeAzi10}. Note
that the above results belong to the range of values of the quota for
qualified majority, which are used in practice or recommended by experts.

\subsection{Large number of member states and a statistical approximation}

Further investigation has confirmed that the existence of such a critical
point is not restricted to the concrete distribution of the population in
European Union. On the contrary, it was reported for a model union containing
$M$ states with a random distribution of population
\cite{SloZyc04,ChaChuMac06,SloZycAPP06}. However, it seems unlikely that we
can obtain an analytical expression for the optimal quota in such a general
case. If the number of member states is large enough one may assume that the
distribution of the sum of the weights is approximately Gaussian
\cite{Owe75,FLMR07,SloZyc07}. Such an assumption allowed us to derive an
explicit approximate formula for the optimal quota for the Penrose square root
voting system \cite{SloZyc07}
\begin{equation}
q_{\mathrm{n}}\ :=\ \frac{1}{2}\left(  1+\frac{\sqrt{\sum_{i=1}^{M}N_{i}}%
}{\sum_{i=1}^{M}\sqrt{N_{i}}}\,\right)  \text{ ,}\label{optiq1}%
\end{equation}
where $N_{i}$ denotes the population of the $i$--th state. In practice it
occurs that already for $M=25$ this approximation works fine and in the case
of the EU-$25$ gives the optimal quota with an accuracy much better than one
percent. Although the value of the optimal quota changes with $M$, the
efficiency of the system, measured by the probability of forming the winning
coalition, does not decrease if the union is enlarged. It was shown in
\cite{SloZyc07} that, according to the central limit theorem, the efficiency
of this system tends to approximately $15.9\%$ if $M\rightarrow\infty$.

It is not difficult to prove that for any fixed $M$ the above expression
attains its minimum if the population of each member state is the same,
$N_{i}=const\left(  i\right)  $. In this way one obtains a lower bound for the
optimal quota as a function of the number of states \cite{SloZyc07}:%
\begin{equation}
q_{\mathrm{min}}\ :=\ \frac{1}{2}\left(  1+\frac{1}{\sqrt{M}}\right)  \text{
.}\label{optbound}%
\end{equation}
Note that the above bound decreases with the number of the states forming the
union as $1/\sqrt{M}$ to $50\%$. Such a behavior, reported in numerical
analysis of the problem \cite{SloZyc04,ChaChuMac06,SloZycAPP06} is consistent
with the so-called Penrose limit theorem -- see Lindner and Machover
\cite{LinMac04}.

\subsection{Optimal quota averaged over an ensemble of random states}

Concrete values of the optimal quota obtained by finding numerically the
minimum of the discrepancy (\ref{Delta}) for the EU-$25$ and the EU-$27$
\cite{SloZyc04,SloZycAPP06,SloZyc10} are consistent, with an accuracy up to
two per cent, with the data obtained numerically by averaging over a sample of
random distribution of the populations of a fictitious union. This observation
suggests that one can derive analytically an approximate formula for the
optimal quota by averaging the explicit expression (\ref{optiq1}) over an
ensemble of random populations $N_{i}$.

To perform such a task let us denote by $x_{i}$ the relative population of a
given state, $x_{i}=N_{i}/\sum_{i=1}^{M}N_{i}$. Since $\sqrt{N_{i}}/\sqrt
{\sum_{i=1}^{M}N_{i}}=\sqrt{x_{i}}$ one can rewrite expression (\ref{optiq1})
in the new variables to obtain
\begin{equation}
q_{\mathrm{n}}(\overrightarrow{x})\ =\ \frac{1}{2}\left(  1+\frac{1}%
{\sum_{i=1}^{M}\sqrt{N_{i}}/\sqrt{\sum_{i=1}^{M}N_{i}}}\,\right)  =\frac{1}%
{2}\left(  1+\frac{1}{\sum_{i=1}^{M}\sqrt{x_{i}}}\,\right)  \ \text{.}
\label{optiq2}%
\end{equation}

By construction, $\overrightarrow{x}=\left(  x_{1},\dots,x_{M}\right)  $ forms
a probability vector with $x_{i}\geq0$ and $\sum_{i=1}^{M}x_{i}=1$. Hence the
entire distribution of the population of the union is characterized by the
$M$-point probability vector $\overrightarrow{x}$, which lives in an $(M-1)$
dimensional simplex $\Delta_{M}$. Without any additional knowledge about this
vector we can assume that it is distributed uniformly on the simplex,
\begin{equation}
P_{D}(x_{1},\dots,x_{M})\ =\ \frac{1}{(M-1)!}\ \delta\left(  1-\sum_{i=1}%
^{M}x_{i}\right)  \text{ .}\label{flat}%
\end{equation}
Technically it is a particular case of the \textsl{Dirichlet distribution},
written $P_{D}(\overrightarrow{x})$, with the Dirichlet parameter set to unity.

In order to get a concrete result one should then average expression
(\ref{optiq2}) with the flat probability distribution (\ref{flat}). Result of
such a calculation can be roughly approximated by substituting $M$-fold mean
value over the Dirichlet measure, $M\langle\sqrt{x}\rangle_{D}$, instead of
the sum into the denominator of the correction term in (\ref{optiq2}),
\begin{equation}
q_{\mathrm{av}}(M)\ :=\ \langle q_{\mathrm{n}}\rangle_{D}\ \approx\ \frac
{1}{2}\left(  1+\frac{1}{M\langle\sqrt{x}\rangle_{D}}\,\right)  \ \text{.}%
\label{optiq3}%
\end{equation}

The mean square root of a component of the vector $\overrightarrow{x}$ is
given by an integral with respect to the Dirichlet distribution
\begin{equation}
\langle\sqrt{x}\rangle_{D}\ =\ \int\limits_{\Delta_{M}}\sqrt{x_{1}}%
\,P_{D}(x_{1},\ldots,x_{M})\,dx_{1}\cdots dx_{M}\text{ .}\label{mean}%
\end{equation}

Instead of evaluating this integral directly, we shall rely on some simple
fact from the physical literature. It is well known that the distribution of
the squared absolute values of an expansion of a random state in an
$M$-dimensional complex Hilbert space is given just by the flat Dirichlet
distribution (see e.g. \cite{BenZyc06}). In general, all moments of such a
distribution where computed by Jones in \cite{Jon91}. The average square root
is obtained by taking his expression (26) and setting $d=M$, $l=1$, $\nu=2$
and $\beta=1/2$. This gives the required average
\begin{equation}
\langle\sqrt{x}\rangle_{D}\ =\ \frac{\Gamma(M)\,\Gamma(3/2)}{\Gamma
(M+1/2)}\ \sim\ \frac{\sqrt{\pi}}{2\sqrt{M}}\text{ .}\label{mean2}%
\end{equation}
Here $\Gamma$ denotes the Euler gamma function and the last step follows from
its Stirling approximation. Substituting the average $\langle\sqrt{x}%
\rangle_{D}$ into (\ref{optiq3}) we arrive at a compact expression
\begin{equation}
q_{\mathrm{av}}(M)\approx\ \frac{1}{2}+\frac{1}{\sqrt{\pi M}}\ =\ \frac{1}%
{2}\left(  1+\frac{2}{\sqrt{\pi}}\frac{1}{\sqrt{M}}\right)  \ \text{.}%
\label{main}%
\end{equation}
This approximate formula for the mean optimal quota for the Penrose voting
system in a union of $M$ random states constitutes the central result of this
work. Note that this expression is averaged over all possible distributions of
populations in the union, so it depends only on the size $M$ of the union and
on the form of averaging. The formula has a similar structure as the lower
bound (\ref{optbound}), but the correction term is enhanced by the factor
$2/\sqrt{\pi}\approx1.128$. In some analogy to the famous \textsl{Buffon's
needle (or noodle) problem} \cite{Ram69}, the final result contains the number $\pi$ -- it
appears in (\ref{main}) as a consequence of using the normal approximation.
The key advantage of the result (\ref{main}) is due to its simplicity.
Therefore, it can be useful in a practical case, if the size $M$ of the voting
body is fixed, but the weights of the voters (e.g. the populations in the EU) vary.

\bigskip

\begin{table}[th]
\label{tab2}
\par
\smallskip
\par
{\renewcommand{\arraystretch}{1.45} }
\par
\begin{center}%
\begin{tabular}
[c]{||c||c|c|c|c|c|c||}\hline\hline
$M$ & $25$ & $27$ & $28$ & $29$ & $...$ & $M\to\infty$\\\hline
$q_{n}[\%]$ & $62.16$ & $61.58$ & $61.38$ & $61.32$ & $...$ & $50.0$\\\hline
$q_{\mathrm{av}}[\%]$ & $61.28$ & $60.86$ & $60.66$ & $60.48$ & $...$ &
$50.0$\\\hline
$q_{\mathrm{min}}[\%]$ & $60.00$ & $59.62$ & $59.45$ & $59.28$ & $...$ &
$50.0$\\\hline\hline
\end{tabular}
\end{center}
\end{table}

Table 2. Optimal quota $q_{n}$ for the Council of the European Union of $M$
member states compared with predictions $q_{\mathrm{av}}$ of the approximate
formula (\ref{main}) and the lower bound $q_{\mathrm{min}}$ given in
(\ref{optbound}). The calculations of the optimal quotas for the EU were based
upon the Eurostat data on the distribution of population for the EU--$25$
(2004) and the EU--$27$ (2010). The extended variant EU-$28$ contains EU-$27$
and Croatia, while EU-$29$ includes also Iceland.

\medskip

\section{Concluding remarks}

In this work we review various arguments leading to the weighted voting system
based upon the square root law of Penrose. However, the key result consists in
an approximate formula for the mean optimal threshold of the qualified
majority. It depends only on the number $M$ of the states in the union, since
the actual distribution of the population is averaged out.

Making use of this result we are in a position to propose a simplified voting
system. The system consists of a single criterion only and is determined by
the following two rules:

\medskip(1) Each member of the voting body of size $M$ is attributed his
voting weight proportional to the square root of the population he represents.

\smallskip(2) The decision of the voting body is taken if the sum of the
weights of members of a coalition exceeds the critical quota $q=1/2+1/\sqrt
{\pi M}$.

\medskip This voting system is based on a single criterion. Furthermore, the
quota depends on the number of players only, but not on the particular
distribution of weights of the individual players. This feature can be
considered as an advantage in a realistic case, if the distribution of the
population changes in time. The system proposed is objective and it cannot a
priori handicap a given member of the voting body. The quota for qualified
majority is considerably larger than $50\%$ for any size of the voting body of
a practical interest. Thus the voting system is moderately conservative, as it
should be. If the distribution of the population is known and one may assume
that it is invariant in time, one may use a modified rule (2\textbf{')} and
set the optimal quota according to the more precise formula (\ref{optiq1}).

Furthermore, the system is transparent: the voting power of each member of the
voting body is up to a high accuracy proportional to his voting weight.
However, as a crucial advantage of the proposed voting system we would like to
emphasize its extendibility: if the size $M$ of the voting body changes, all
one needs to do is to set the voting weights according to the square root law
and adjust the quota $q$ according to the rule (2). Moreover, for a fixed
number of players, the system does not depend on the particular distribution
of weights. This feature is specially relevant for voting bodies in corporate
management for which the voting weights may vary frequently.

It is our pleasure to thank W.~Kirsch, M.~Machover, F.~Pukelsheim for fruitful
discussions and to E.~Ratzer for helpful correspondence. We are obliged to the
anonymous referee for numerous comments and suggestions which allowed us to improve the
article. Financial support by the European grant COCOS is gratefully acknowledged.

\end{document}